\documentclass{elsart}
\usepackage{graphicx,amssymb}
\begin{document}
\begin{frontmatter}
\hfill{\bf BARI-TH 483/04}\\
\title{Deterministic Annealing as a jet clustering algorithm in hadronic collisions}
\author{L. Angelini$^{a,b,c}$, G. Nardulli$^{a,b,c}$}
\author{ L. Nitti$^{b,c,d}$, M. Pellicoro$^{a,b,c}$, D. Perrino$^{a,c}$ and S. Stramaglia$^{a,b,c}$}
\address{$^{a}$Dipartimento Interateneo di Fisica, Bari, Italy}
\address{$^{b}$TIRES, Center of Innovative Technologies for Image Detections and Processing, Bari,
Italy}
\address{$^{c}$I.N.F.N., Sezione di Bari, Italy}
\address{$^{d}$DETO, Universit\`a di Bari, Italy}
 \begin{abstract}
We show that a general purpose clusterization algorithm, Deterministic Annealing, can be
adapted to the problem of jet identification in particle production by high energy collisions.
In particular we consider the problem of jet searching in events generated at hadronic
colliders. Deterministic Annealing is able to reproduce the results obtained by traditional jet
algorithms and to exhibit a higher degree of flexibility.
\end{abstract}
\end{frontmatter}
\section{Introduction}
In high energy hadron-hadron collisions, events with high transverse energy are characterized
by highly collimated particle jets, reflecting hard scattering processes at parton level.
Radiation and pair production processes hide the information on the original partons momenta.
To bridge the gulf between experimental results expressed in terms of hadron properties, and
the theory, whose ingredients are quarks and gluons, a reconstruction processes is needed. By
this process hadrons in the final states are grouped in jets and many dedicated algorithms have
been proposed to this purpose. These algorithms, that shall be reviewed in section \ref{sec2},
appear to be reasonable recipes taking into account geometrical considerations and theoretical
prescriptions. It can be guessed that in this way one is solving an optimization problem,
trying to minimize some cost functions. This is exactly at the basis of the so-called
clustering problem. Here one looks for the optimal partition of a given set of objects in
classes on the ground of some similarity property. This task is performed minimizing a
prescribed cost function that is to be adapted to the problem under investigation. In a recent
paper \cite{barijet02} it has been shown that a particular clustering algorithm, the so-called
Deterministic Annealing (DA) \cite{RoGuFo90,RoGuFo93,Rose98}, can be adapted to the study of
the hadronic jets in high energy $e^+e^-$ scattering. Essentially, DA can give the same results
of the standard Durham algorithm in a faster way, as a consequence of a lower computational
complexity. In this work we try to extend the use of DA to hadron-hadron collisions taking into
account the peculiarities of the jet production in this type of interaction. In particular in
this kind of interactions only a part of the particles in the final state can be associated to
partons coming from a hard scattering process.
\par Deterministic Annealing, in a version that allows data analysis in terms of a number of
clusters either fixed or variable, will be presented in section \ref{sec3}. In section
\ref{sec4} results from the application of this method to simulated events will be presented
and compared with those obtained by a Cone algorithm. Section \ref{sec5} is dedicated to
comments and conclusions.

\section{Jet clustering algorithms\label{sec2}}
The need to associate energy and momentum of particles in the final state to the four-momentum
of unobservable partons is realized through jet clustering algorithms\footnote{For a review of
these and other jet algorithms see \cite{rassegne2}. For a review of the Montecarlo generators
and their connections with the jet algorithms see \cite{rassegne}.}. The most common of them
can be classified in two categories:
 \begin{itemize}
 \item Association algorithms that use an iterative procedure. For every pair of particles with
 four momentum $p_i$ and $p_j$, a test variable $y_{i,j}=f(p_i,p_j)$ is calculated.
 This test variable is then compared to a given threshold parameter $y_{cut}$
and the pair is recombined into a new pseudo-particle $k$ of four-momentum $p_k=p_i+p_j$ (E
scheme, but other schemes have also been considered) provided that $y_{ij}\,\leq \,y_{cut}$.
The algorithm is then reiterated to the new set of (pseudo)particles and it stops when, for all
pairs, $y_{ij}\,\geq \,y_{cut}$. The number of pseudoparticles at the end of the algorithm
counts the number of jets, which is therefore fixed by $y_{cut}$. The ancestor of these jet
algorithms is the JADE algorithm \cite{JADE1,JADE2} where the jet resolution variable is
defined as
\begin{equation}\label{jadeyij}
  y_{ij}=y^J_{ij}\equiv \frac{2E_iE_j(1-\cos\theta_{ij})}{E^2_{vis}}\ ,
\end{equation}
where $E_{vis}$ is the visible energy, i.e. the sum of energies for all particles observed in
the final state, $E_i$, $E_j$ are the particles energies, and $\theta_{ij}$ their angular
separation. The theoretical advantage of this recombination scheme lies in the absence of
collinear and infrared singularities, as the regions of phase space where these divergences
could be generated are automatically excluded. However it is clear that also particles at very
different angles can be recombined in one pseudo-particle, and this fact can give rise to the
appearance of \emph{ghost} jets along directions where no particles are present. This problem
suggested to modify the test variable in the following way (Durham algorithm
\cite{Durham1,Durham2,Durham3})
\begin{equation}\label{durhamyij}
  y_{ij}=y^D_{ij}\equiv\frac{2\,min\{E_i^2,E_j^2\}(1-\cos\theta_{ij})}{E^2_{vis}}\ .
\end{equation}
Successively yet another variable has been introduced $v_{ij}=2(1-\cos\theta_{ij})$. Firstly
the pairs of particles are ordered following this variable, then the precedent scheme is
applied. If the recombination fails ($y_{ij}\,\ge \,y_{cut}$, the softest (pseudo-)particle is
freezed and hindered from being an attractor for other particles. This mechanism avoids soft
collinear particles to be the seed for unwanted jets. The algorithm that implements these new
rules is known as Cambridge algorithm \cite{Cambridge}.
\item To a second class belong algorithms that associate particles in a jet only on the ground of
geometrical properties. The prototype for them is the Cone algorithm defined in the Snowmass
Convention \cite{Huth}. Here in the first step the few particles having a transverse energy
$E_T$ greater than a fixed threshold $E_T^0$ are selected as \emph{seeds} for jets.
Subsequently the particles lying in a cone of given radius $R_0$ in the pseudorapidity-azimuth
plane around each seed are associated with a jet, whose direction is fixed by an iterative
procedure. More refined approaches consider the possibility of recombination and splitting of
these proto-jets.
\end{itemize}
Here we stress an important difference between these two categories. While for the algorithms
of the first kind jets include all the particles and their number can be fixed a priori, for
the algorithms of the second kind the number of jets is essentially determined by the number of
particles used as seed and a varying part of particles is excluded from the classification.
This is the reason why the former scheme is used in the case of electron-positron scattering
and the latter in the case of hadronic diffusions, where not all the particles are produced in
hard interactions.
\section{Deterministic Annealing\label{sec3}}
As we said the clustering problem consists of the optimal grouping of a set of data points so
that points in the same class are more similar than points in different classes. Deterministic
Annealing is inspired by an analogy to the annealing procedure that consists of maintaining a
system at thermal equilibrium while gradually lowering the temperature. The process assures
that, in the limit of low temperature, the global free energy minimum is attained. The word
\emph{deterministic} refers to the fact that, as we shall see, thermal equilibrium is obtained
minimizing directly the free energy, in opposition to the stochastic simulation used by
Simulated Annealing \cite{SA}. We introduce here a formulation of DA called Mass-Constrained
Clustering (MCC) \cite{RoGuFo93,Rose98} that is particularly suitable for our application. In
effect in this formulation the number of clusters is not fixed a priori, as it happened in the
precedent application of DA to the jet searching problem \cite{barijet02}, but is the result of
the calculation.

\begin{figure}[ht]
\begin{center}
\includegraphics[width=7.5cm]{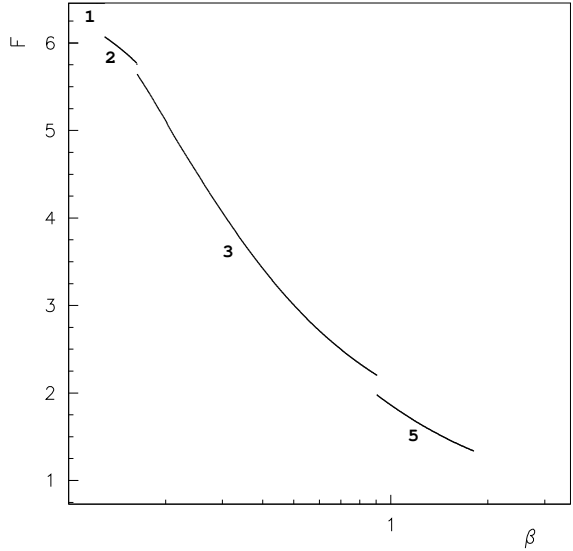}
\end{center}
\caption{The phase diagram for a simulated p-p scattering event (see sect. \ref{sec4})}
\label{phase}
\end{figure}

Let us consider two sets, the set of the data points $x\in X$ we want to classify and the set
of the vectors representative of the clusters $y\in Y$, also called \emph{code-vectors}. The
MCC approach introduces an infinite number of code-vectors; at each stage of the annealing
process only a limited portion of them are distinct, so one introduces a quantity $p_i$
denoting the fraction of code-vectors which are coincident and represent the same cluster $i$.
One defines also the local distortion $d(x,y_i)$ between each data point $x$ and each effective
code-vector $y_i$. The global distortion $D$ is defined as
\begin{equation}\label{dist}
  D=\sum_x\sum_i p(x,y_i)d(x,y_i)=\sum_x p(x)\sum_i p(y_i|x)d(x,y_i)\ ,
\end{equation}
where $p(x,y)$ is the joint probability distribution, $p(x)$ is the probability of each data
set element and $p(y_i|x)$ is the conditional probability relating the element $x$ with
code-vector $y_i$, i. e. the probability to associate $x$ with cluster $i$. Following the
analogy with a statistical physics system, $D$ plays the role of the internal energy which, in
the limit of zero temperature, one wants to minimize. In this limit one obtains the hard
clustering solution, in which the association probabilities are zero or one. At finite
temperature the minimum of the Helmholtz free energy $F$ determines the distribution at thermal
equilibrium. This minimum is given by :
\begin{equation}\label{fren}
  F^*=-T\sum_x\ln\, Z_x\ ,
\end{equation}
where $Z_x$ is the partition function for the single data point
\begin{equation}\label{Zx}
  Z_x=\sum_i p_i\,e^{-d(x,y_i)/T}\,.
\end{equation}
As a consequence the conditional probabilities are given by the Gibbs distribution
\begin{equation}\label{Gibbs}
  p(y_i|x)=\frac{p_i\,e^{-d(x, y_i)/T}}{Z_x} \,.
\end{equation}
Imposing the free energy minimization under the constraint $\sum_i p_i=1$, one obtains that the
optimal set of code-vectors $\{y_i\}$ must satisfy the equations
\begin{equation}\label{gradeq}
  \sum_x p(x)\,p(y_i|x) \nabla_{y_i}d(x,y_i)=0\,,
\end{equation}
while
\begin{equation}\label{py}
  p_i=\sum_x p(x) p(y_i|x)=p(y_i)\,.
\end{equation}
From eq. (\ref{gradeq}) one obtains that the positions of the code-vectors are determined, for
a squared error distortion $d(x,y_i)=|x-y_i|^2$, by
\begin{equation}\label{yi}
  y_i=\frac{\sum_x x p(x) p(y_i|x)}{p(y_i)}\ .
\end{equation}
 The annealing process starts at high temperature. From
(\ref{Gibbs}) it is clear that the association probabilities are uniform, the system is
completely disordered and the code-vector set collapses to a single point. This unique
code-vector has $p(y_1)=1$, every point is associated with this code-vector with probability
$1$, $p(y_1|x)=1$, and equation (\ref{yi}) gives the position of the centroid of the data set
$y_1=\sum_x p(x)\,x$. During the cooling process one encounters phase transitions which consist
of an increase in the number of code-vectors through a sequence of cluster splittings. The
temperature plays the role of the resolution parameter at which the data set is clustered and a
complete hierarchical clustering can be obtained up to the extreme situation at zero
temperature when there is a code-vector for each point of the data set. This process is
described in Fig.(\ref{phase}) where the behavior of the Free energy $F$ as a function of
$\beta=1/T$ is shown for a typical event among those analyzed in the next section.
From a practical point of view, Mass Constrained Clustering can be implemented by an algorithm
that here we briefly sketch. Starting from a low value of $\beta$ one introduces two clusters
with coordinates slightly perturbed with respect to the centroid coordinates and equal
probability for every point to be associated with each cluster. Then one minimizes the free
energy iterating the equations:
\begin{equation}\label{MCC1}
  p(y_i)=\sum_x p(x) p(y_i|x)
\end{equation}
\begin{equation}\label{MCC2}
  p(y_i|x)=\frac{p(y_i)\,e^{-\beta d(x, y_i)}}{\sum_j p(y_j)\,e^{-\beta d(x,y_j)}}
\end{equation}
\begin{equation}\label{MCC3}
  y_i=\frac{\sum_x x p(x) p(y_i|x)}{p(y_i)}
\end{equation}
until one finds convergence in $y_i$. If $\beta$ is low enough, it comes out that these two
clusters are coincident. The next step is to cool the system, $\beta \rightarrow \alpha\beta$,
always iterating equations (\ref{MCC1},\ref{MCC2},\ref{MCC3}) until a solution corresponding to
two different code-vectors (the first phase transition) is encountered. Subsequently one goes
on by introducing, for each code-vector location, two perturbed code-vectors which share the
association probability of each data point, raising $\beta$ and determining the new
code-vectors coordinates. Each pair of code-vectors will be merged until a critical $\beta$ is
reached, in which case one of the pairs will originate two effective code-vectors. The process
will be stopped when a sufficient resolution ($\beta$ value or number of clusters) is reached.

In order to apply apply this algorithm to the problem of jet search in hadronic collisions we
must first choose a distortion measure. We considered the squared error distortion in the
pseudorapidity-azimuth plane
\begin{equation}\label{dist2}
  d(x,y_i)=(\eta_x-\eta_i)^2+(\phi_x-\phi_i)^2\,.
\end{equation}
The other ingredient is the weight $p(x)$ to assign to each particle. As our purpose was to
make a comparison with the Cone algorithm, we assigned to a particle $x$ with transverse energy
$E_T$ the weight
\begin{equation}\label{p(x)}
  p(x)=\frac{E_T^x}{\sum_x E_T^x}\ .
\end{equation}
This assignment, together with (\ref{dist2}), has the interesting property that the coordinates
of a jet, as defined by the Cone algorithm,
\begin{equation}\label{jet_coor}
  \eta^J=\frac{1}{E_T^J}\,\sum_{k\in J}\,E_T^k \eta^k \qquad
  \phi^J=\frac{1}{E_T^J}\,\sum_{k\in J}\,E_T^k \phi^k \qquad
  E_T^J=\sum_{k\in J}\,E_T^k \ ,
\end{equation}
are exactly recovered by the DA algorithm in the limit of hard clustering ($\beta \to \infty$).
In this limit, indeed, the association probabilities of each data point (particle) to a cluster
(jet) in eq. (\ref{MCC2}) become $0$ or $1$ and from eq. (\ref{MCC3}) one obtains exactly eqs.
(\ref{jet_coor}).
\section{Results and discussions\label{sec4}}
We are now in position to explore the possibility of applying the Mass Constrained Clustering
version of Deterministic Annealing to the problem of jet search in hadronic colliders. To this
purpose we generated 2000 events from proton-proton scattering at $14\ TeV$ by the PYTHIA
\cite{Pyt1,Pyt2} Monte Carlo generator; a bias in the transverse energy $E_T$ of the initial
partons was introduced, corresponding to $E_T=100\ GeV$  for 1000 events (sample A) and
$E_T=200\ GeV$ for the other 1000 events (sample B); initial and final state radiation was
allowed. With this bias, a clear back-to-back two jet structure is expected. Results from
application of DA where systematically compared with those obtained by the Cone algorithm
described in section \ref{sec2}: the Cone algorithm parameters, the transverse energy threshold
$E_T^0$ and the cone radius $R_0$ have been fixed to $2\ GeV$ and $0.7$, respectively.

\begin{figure}[ht]
\begin{center}
\includegraphics[width=7.5cm]{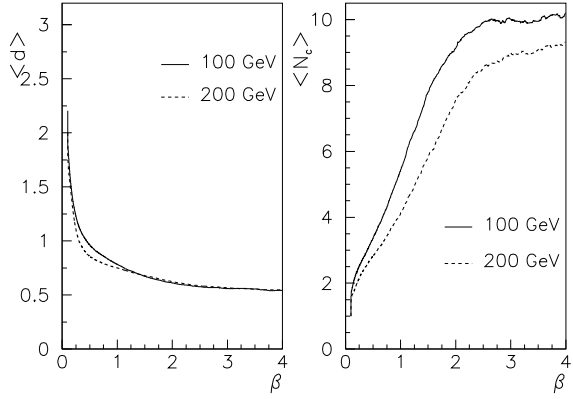}
\end{center}
\caption{First results from the application of DA algorithm to simulated events. Left side:
$\langle d \rangle$, averaged over all the events, vs $\beta$. Right side: the mean clusters
number $\langle N_c\rangle$, vs $\beta$.} \label{dncvsbeta}
\end{figure}

  We calculated first two quantities that can be easily used for a comparison with the Cone algorithm.
The first quantity is the mean distance of each particle from a code vector $j$, defined as
\begin{equation}\label{mdist}
  \langle d \rangle =\frac{1}{N_c}\sum_{j=1}^{N_c}\frac{1}{p(y_j)}\,\sum_x \sqrt{d(x,y_j)}\,
  p(x)\,
  p(y_j|x)\,,
\end{equation}
where $N_c$ is the number of clusters found. $\langle d \rangle$ is a decreasing function of
$\beta$ attaining its maximum value at $\beta=0$, when there is only one cluster, and its
minimum value, that is zero, at $\beta=\infty$ when every particle is a cluster by itself. This
quantity, averaged over all the events, is shown, as a function of $\beta$ in the left part of
fig. \ref{dncvsbeta}.
We can see that there is no practical difference between the two analyzed samples: in either
case $\langle d \rangle$ decreases quickly for low values of $\beta$, due to the growth in the
number of clusters, then the descent becomes very slow. This behavior is the signal that the
particle distribution in the events we are analyzing is such that the initial partition in few
clusters is preserved when $\beta$ is increased, apart from fragments of low weight. This
robustness is confirmed by the second quantity we calculated, that is the mean number of
clusters $\langle N_c\rangle$, whose behavior, as a function of $\beta$, is shown in the right
part of fig. \ref{dncvsbeta}. We see that the region of extreme fragmentation, that ends the
clustering process, is far away at $\beta=4$.

\begin{figure}[ht]
\begin{center}
\includegraphics[width=7.5cm]{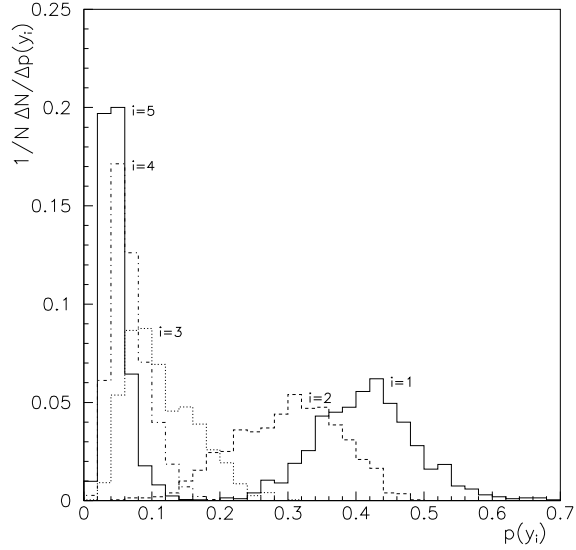}
\end{center}
\caption{The probability distributions for the five most probable clusters at $\beta=1.4$.}
\label{prob5}
\end{figure}

\begin{figure}[ht]
\begin{center}
\includegraphics[width=7.5cm]{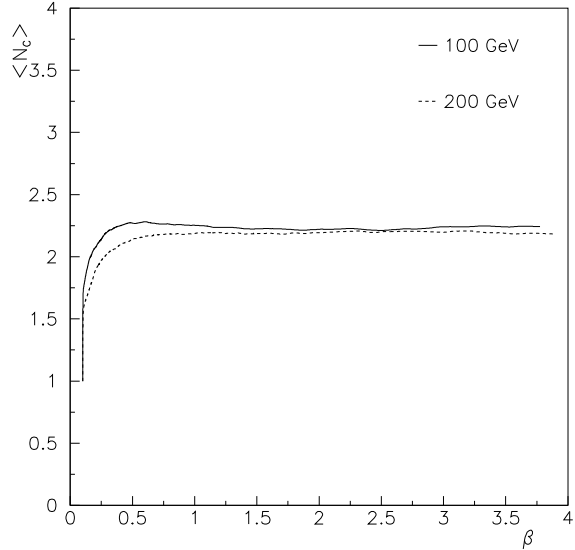}
\end{center}
\caption{The mean clusters number $\langle N_c\rangle$, vs $\beta$. Only clusters having
probability greater than $p_0=0.15$ have been considered.} \label{ncvsbeta}
\end{figure}

 How to determine the two jet nature of our events? To answer this question we note that
the DA recipe cannot be yet considered complete, because we still have two problems. The first
problem is that the annealing process must be stopped at some $\beta$ value to avoid the
extreme cluster fragmentation produced by the $\beta \to \infty$ limit. The second problem
arises because only part of the clusters can be attributed to the scattered partons. Therefore,
once we choose $\beta$, we need also a criterion to select real jets from clusters. In the Cone
algorithm these questions are controlled, as mentioned before, through two parameters: the
transverse energy threshold $E_T^0$ and the cone radius $R_0$. We remember that DA introduces a
probability measure for the clusters, the expression (\ref{MCC1}). A peculiarity of these
probabilities is that the two jet nature of the events here analyzed produces, in the $\beta$
region where $\langle d \rangle$ has a smooth behavior ($\beta \gtrsim 1$), two clusters of
high probability, while to the remaining clusters only a small fraction of unity is assigned.
To illustrate this feature the probability distributions for the five most probable clusters at
$\beta=1.4$ are shown in fig. \ref{prob5} for the events from sample A.
A small value of cluster probability reflects the fact that the particles assigned to this
cluster with a good association probability are few and have a small weight (transverse
energy). So it is natural to consider jets only the clusters that survive a cut in the
probability value. For example, we see in fig. \ref{ncvsbeta} how a threshold at $p_0=0.15$
influences the $\beta$ dependence of the mean number of clusters $N_c$. Now this quantity goes
rapidly to a value close to $2$, i.e. the expected value for our sample.

At this point we are ready to illustrate how DA is able to reproduce the results obtained by
the Cone algorithm. We performed the annealing process up to a $\beta$ value of $1.4$ and
accepted only clusters with probability greater than $p_0=0.025$. With these values we obtained
$\langle d\rangle = 0.69 \pm 0.11$, close to the value of $R_0=0.7$ used for the Cone
algorithm. This could be expected because $\beta$ has an effect on the association probability
of a particle to a cluster (see (\ref{MCC2})) that is comparable to that of the parameter $R_0$
for the Cone algorithm, if one puts $\beta\sim 1/2R_0^2$. No fine tuning of these parameters
was performed, because this is not the aim of this article.

\begin{figure}[ht]
\begin{center}
\includegraphics[width=7.5cm]{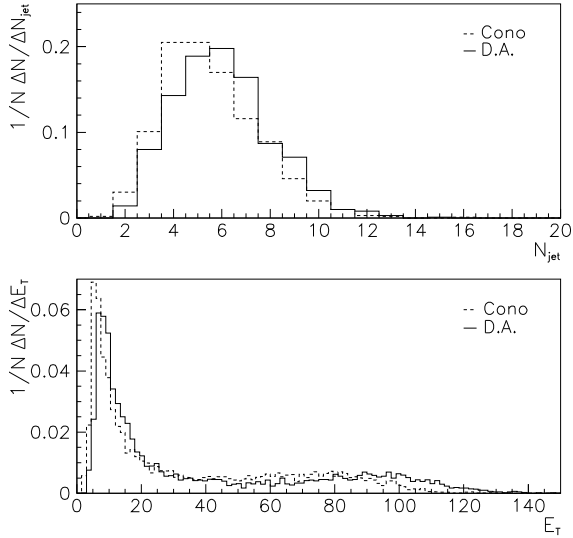}
\end{center}
\caption{The clusters number distribution (top) and the clusters transverse energy distribution
(bottom) for the DA algorithm (solid line) and the Cone algorithm (dashed line). (Events are
from sample A.)} \label{confr1}
\end{figure}

The comparison between the two algorithms is reported in fig. \ref{confr1} for two observables:
the number of clusters and their transverse energy distribution. Some differences can be noted,
in particular there is a more pronounced tail in the $E_T$ distribution for the DA algorithm.
This can be easily explained by the fact that, assuming $\langle d\rangle \sim R_0$, $R_0$ is a
sharp threshold for the Cone algorithm, while for the DA algorithm $\langle d\rangle$ is the
mean cluster radius.
Another minor discrepancy is in the $N_c$-distribution that, for the DA algorithm, is slightly
shifted to higher values of $N_c$. We could get rid of these differences modifying the values
of $\beta$ and $p_0$, but, as we said, we found this job useless, not least because we used a
very simple Cone algorithm, where, for example, no recombination or splitting mechanism for
proto-jets have been considered.
\begin{figure}[ht]
\begin{center}
\includegraphics[width=7.5cm]{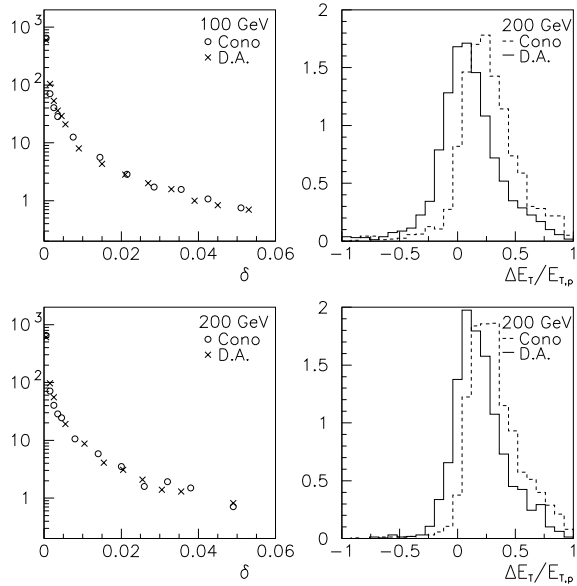}
\end{center}
\caption{The distribution of the angular distance between partons and jets (left) and the
clusters transverse energy distribution (right) for the DA algorithm (solid line) and the Cone
algorithm (dashed line).} \label{confr2}
\end{figure}
A more interesting question to ask is which algorithm better reproduces the properties of the
partons originating the jets. To this purpose we introduced two variables for each parton
participating in the hard initial scattering and for the cluster nearest to it in direction.
The first variable describes the ability to identify the parton direction:
\begin{equation}\label{delta}
  \delta=\frac{1-\cos \alpha}{2}
\end{equation}
where $\alpha$ is the angular separation between parton and jet axis. The other quantity
measures the ability to trace the transverse energy of the parton:
\begin{equation}\label{deltaet}
  \Delta=\frac{E_{T,p}-E_{T,c}}{E_{T,p}}
\end{equation}
where $E_{T,p}$ and $E_{T,c}$ are the transverse energies of the parton and the jet
respectively. Their distributions for the two algorithms and the two data samples are shown in
fig. \ref{confr2}. We can see that, while for the $\delta$ distribution there are no practical
differences between the two algorithms, DA seems to be more efficient in recovering the hard
parton transverse energy.
\section{Conclusions\label{sec5}}
We have compared the results found by the Cone algorithm with those obtained by a clustering
algorithm based on the Deterministic Annealing procedure. The latter has been adapted to the
process studied in this article, i.e. jet identification in particle production by high energy
hadronic collisions, by introducing a suitable distortion measure and using temperature and
cluster probability as parameters. Other choices are possible. For example one could take into
account that the phase transition producing the splitting of a cluster occurs at a temperature
proportional to the variance of the cluster itself \cite{Rose98}. So a characteristic of well
defined clusters is that they are stable for a wide range of temperature and this stability
property could be used in jet recognition.
 \par From this preliminary analysis we cannot conclude that the DA algorithm should be preferred
to the Cone algorithm, even if the good results for the number of clusters (\ref{ncvsbeta}) and
on the parton transverse energy should not be neglected. In any case we think that the
jet-physics community should consider DA as a possible and serious alternative. The use of a
geometrical definition of jet appear indeed too simplifying with respect to the theoretical
descriptions. On the other hand, the DA algorithm looks at the properties of the density
distribution in the momentum space, which is the reason why the recombination and splitting
mechanisms are automatically incorporated.
\par Moreover, there is another general question that could be solved in this calculation scheme.
It is true that the choice of jet definition is a matter of convention and that the important
thing is to use the same definition in theoretical predictions and in experimental analysis.
However it cannot be considered satisfactory that, while there is a unique theory explaining
jet production and properties, different definitions are used in hadronic and in leptonic
collisions. The purpose of this paper is to demonstrate that this difficulty could be overcome
using the same algorithm, so that one can focus all the efforts in the most important question,
i. e. the similarity property used to decide if two particles should be assigned to the same
jet. Using a correct definition of this quantity, indeed, one can take into account important
theoretical peculiarities, as infrared and collinear safety or formation of "ghost" and "junk"
jets \cite{Seym94,BeMe98}. These kinds of similarity measure have been used, until now, only
for $e^+-e^-$ collisions and embodied in algorithms with poor performance, since they have to
loop on all the particles' pairs. We hope to have clarified (see also \cite{barijet02}) that
they could be used for any kind of interaction, without giving up the reduced computational
complexity that geometrical algorithms share with the method we propose.
\par\noindent{\bf Acknowledgements}
\par\noindent
We thank D. Di Bari for most useful discussions and suggestions.


\begin{thebibliography}{10}

\bibitem{barijet02}
L.~Angelini, P.~De Felice, M.~Maggi, G.~Nardulli, M.~Pellicoro, and
  S.~Stramaglia.
\newblock {\em Physics Letters B}, 545:315--322, 2002.

\bibitem{RoGuFo90}
K.~Rose, E.~Gurewitz, and G.~C. Fox.
\newblock {\em Pattern Rec. Letters}, 11:589, 1990.

\bibitem{RoGuFo93}
K.~Rose, E.~Gurewitz, and G.~C. Fox.
\newblock {\em IEEE Trans. Pattern anal. Machine Intell.}, 15:785--794, 1993.

\bibitem{Rose98}
K.~Rose.
\newblock {\em Proc. of the IEEE}, 86:2210--2239, 1998.

\bibitem{rassegne2}
S.~Moretti, L.~Lonnblad, and T.~Sjostrand.
\newblock {\em JHEP}, 9808:001, 1998.

\bibitem{rassegne}
A.~Ballestrero et~al.
\newblock Report of the qcd working group.
\newblock Technical Report hep-ph/0006259.

\bibitem{JADE1}
W.~Bartel et~al. [JADE~Collaboration].
\newblock {\em Z. Phys. C}, 33:23, 1986.

\bibitem{JADE2}
S.~Bethke et~al. [JADE~Collaboration].
\newblock {\em Phys. Lett. B}, 213:235, 1988.

\bibitem{Durham1}
Yu.~L. Dokshitzer.
\newblock in proc. workshop on jet studies at lep and hera, durham, 1990.
\newblock {\em J. Phys. G}, 17:1572ff, 1991.

\bibitem{Durham2}
S.~Catani, Yu.~L. Dokshitzer, M.~Olsson, G.~Turnock, and B.~R. Webber.
\newblock {\em Phys. Lett. B}, 269:432, 1991.

\bibitem{Durham3}
N.~Brown and W.J. Stirling.
\newblock {\em Z. Phys. C}, 53:629, 1992.

\bibitem{Cambridge}
Yu.~L. Dokshitzer, G.~Leder, S.~Moretti, and B.~Webbe.
\newblock {\em JHEP}, 9708:001, 1997.

\bibitem{Huth}
J.~Huth et~al.
\newblock In E.L. Berger, editor, {\em Proceedings of the 1990 Division of
  Particles and Fields Summer Study, Snowmass 1990}. World Scientific,
  Singapore, 1992.

\bibitem{SA}
S.~Kirkpatrick, C.D. Gelatt, and M.P. Vecchi.
\newblock {\em Science}, 220:671, 1983.

\bibitem{Pyt1}
T.~Sj\"ostrand.
\newblock {\em Comp. Phys. Commun.}, 82:74, 1994.

\bibitem{Pyt2}
T.~Sj\"ostrand, L.~L\"onnblad, and S.~Mrenna.
\newblock Pythia 6.2 physics and manual.
\newblock Technical Report hep-ph/0108264.

\bibitem{Seym94}
M.~Seymour.
\newblock {\em Z. Phys. C}, 62:127, 1994.

\bibitem{BeMe98}
S.~Bentvelsen and I.~Meyer.
\newblock {\em Eur. Phys. J. C}, 4:623, 1998.

\end{thebibliography}

\end{document}